# Study of Rb - vapor coated cell; atomic diffusion and cell curing process

S.N. Atutov[1], F.A. Benimetskiy[1,2], A.I. Plekhanov[1,2], V.A. Sorokin[1,2]

1. *Institute of Automation and Electrometry of the Siberian Branch of the Russian Academy of Science, Koptyug Ave. 1, Novosibirsk, Russia, 630090*
2. *Novosibirsk State University, Pirogova street 2, Novosibirsk, Russia, 630090*

**Abstract.** We present the results of the study of an optical - resonant cell filled by a vapor of the Rb atoms and coated with a non – stick polydimethylsiloxane (PDMS) polymer. We show that it is possible to define correctly the diffusion coefficient of the atoms in the coating, using geometric parameters of the cell and the vapor density in the cell volume only. The dependence of the diffusion coefficient on the cell curing time is presented. It is shown that the mysterious cell curing process can be explained in terms of the polymerization of the polymer coating by alkali atoms. Anomalous long dwell time of the Rb atoms on the PDMS coating is discussed as well.

## 1. Introduction

There are many publications in the literature devoted to the experiments with optical resonant cells. This kind of cells is widely used for a lot of experiments with the magneto-optical trapping of radioactive isotopes or rare atoms [1,2], atomic clocks [3], magnetometers [4-6], fundamental symmetry studies [7], electromagnetically induced transparency[8], spin squeezing [9], long-lived entanglement [10] and quantum memory [11]. The major difficulty in many resonant cell experiments lies in the atomic vapor interaction with the inner wall of the resonance cell. It is known that when atoms collide with the surface, they undergo an attractive potential which range depends on electronic and atomic structures of the surface and atoms. The fraction of the atoms can be trapped in the attractive potential well on the surface by a physical adsorption process. The adsorbed particles might be desorbed back to the gas phase from the surface after a dwell (sticking) time, if the energy that has been imparted to them from the surface is enough to overcome the surface Van der Waals attractive force. This dwell time is an important characteristics of the atoms – wall interaction for all resonant cell experiments. This is because a long dwell time causes a large loss of the interacted atoms on account of their radioactive decay on the surface that greatly decreases the efficiency of the trapping. In the case of polarized atoms, a long dwell time causes a fast atomic spin relaxation on the cell walls.

The use of the polymer antirelaxation coatings has substantially overcome the difficulties in the above studies. This kind of polymer coatings is characterized by a low surface potential, which greatly reduces the atom – surface interaction time and the probability of absorption of the surface atoms in one collision. For the first time this kind of coatings was suggested in [12] and studied in [13, 14]. Now the literature contains a number of publications devoted to the efficient use of polymer coatings in resonance experiments (see, [15] and references therein). Given the importance of the application of polymer coatings in optical experiments with resonant excitation of atoms, there were made extensive studies of the used polymers (see, [16, 17] and references therein), and a number of important and interesting results was obtained.

The aim of this paper is to study the diffusion of the Rb atoms in the polydimethilsoloxane (PDMS) film in the coated optical – resonant cell. We model the Rb diffusion in the cell and in the coating. We use this model to measure the Rb diffusion coefficient in the coating and its dependence on the cell curing time. We also explain our measurements of the diffusion coefficient in a cuvette filled by a liquid PDMS compound. Finally, we discuss the essence of mysterious curing process which is always observed in the experiments with the resonant cells covered by any polymer coatings. We believe that the results of the experiment can be used to improve the characteristics of a magneto - optical traps, atomic frequency standards, magnetometers etc. and give better understanding details of the processes of the interaction of alkali atoms with a polymer surface.

## 2 Experiment

The main part of the setup (see Fig. 1) is a cell made of a Pyrex glass bulb (1), attached to vacuum ionic and turbo pumps with a vacuum gauge and to an atomic vapor source of natural isotopic mixture of Rb. The cell is connected to the pumps through a glass pump tube (2) with valves (3,4), the source - through a small valve (5) and glass capillary (6). The dimensions of all parts of the cell are carefully measured - the bulb has internal diameter $R = 7$cm, the exit tube internal diameter is $2R_V = 2.1$ cm, length $L_V = 16$ cm and the capillary internal diameter is $2r_s = 0.3$ cm, length $L_s = 4.5$ cm. During the experiments both pumps provide a vacuum of $10^{-10}$ mbar. The inner surface of the cell, pump tube, capillary and parts of valves 4, 5 directed to the cell are covered by the PDMS coating. This coating is prepared from 10 % solution of a commercial PDMS liquid material (Mw.182.600, $4 \cdot 10^3$ cm$^2$s$^{-1}$ viscosity, secondary standard, Aldrich Chemical Company. Inc.) in ether. The chemical formula for the PDMS is CH$_3$[Si(CH$_3$)$_2$O]$n$Si(CH$_3$)$_3$, where $n$ is the number of repeating SiO(CH$_3$)$_2$ monomer units of a molecule chain. The film thickness $L_p$ is determined by interferometric measurements and is found to be equal to 10 μm.

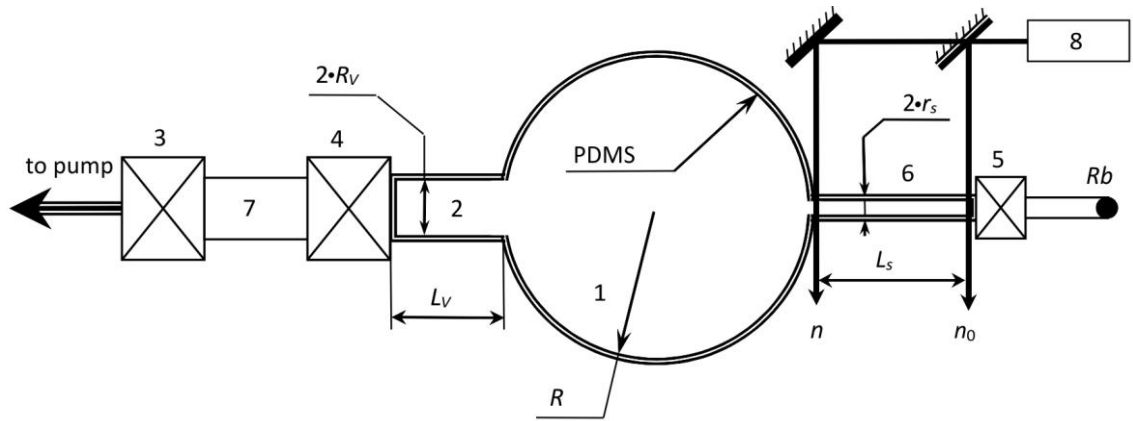

Fig.1. (1) - cell, (2) - exit tube, (3,4,5) - valves, (6) – capillary, (5) - flange, (7) – pump tube, (8) - diode laser

The coating procedure is the following. First, we heat and pump the cell to obtain a residual-gas pressure of $10^{-10}$ mbar. It usually takes several weeks to get good vacuum conditions. Before producing the coating, special care is taken to clean the inner surfaces of the bulb and the pump tube from the traces of chemical active gases, especially from water. It is done by a treatment of these surfaces by ions bombardment in glow discharge of neon at the pressure of 5 mbar. The inner surface of the relatively narrow capillary is cleaned by DC discharge by applying of a high voltage between valves 4,5. We consider the cell to be clean when the discharge luminescence has a bright neon color and it does not change during several hours of discharging. After cleaning the cell is disconnected from the pump keeping valves 4,5 closed, then the cell is set in a vertical position (with the source of rubidium down) and then the pump tube 7 is completely filled by the PDMS solution in ether. Soon after that we open the valve 4 to allow the mixture to come inside the cell and, finally, after rinsing of the cell during several seconds the mixture is poured out of the cell.

The density of the Rb vapor is measured by means of the detection of the intensity of the atomic fluorescence by a photodiode connected to a data acquisition system. The fluorescence is exited by a free-running diode laser (8) with a frequency tuned to a Rb atom resonant transition of 780 nm. The fluorescent signals are processed by a digital oscilloscope connected to the computer. The absolute Rb vapor density at the atomic source is estimated from the temperature of the Rb metal drop [18].

Despite the careful cell cleaning, we found that the freshly coated cell with residual-gas pressure of $10^{-10}$ mbar did not show any fluorescence from the Rb atoms, meaning that the density of the atoms in the cell is very small. We carried out a routine curing procedure using the Rb vapor. Maintaining continuous pumping of the cell, we heated up the source of the Rb atoms so that the pressure of the alkaline vapor in the cell was about $10^{-6}$ mbar. For each measurement, the pressure of the Rb vapor was reduced by the source keeping at room temperature and the vapor density was measured with valve 4 being blocked. The measurement is done at two points in the capillary: one near the valve 5 ($n_0$), the other at the exit of the capillary very close to the bulb (n). Figure 2 shows how the equilibrium density n depends on the duration of the curing process. At the beginning of the curing process, the equilibrium density is very small. After 7 days of continuous curing, the equilibrium density approaches a limit, but still never reaches $n_0$ level. From these measurements, one can deduce that the increase in the equilibrium density after curing is approximately a factor $10^4$.

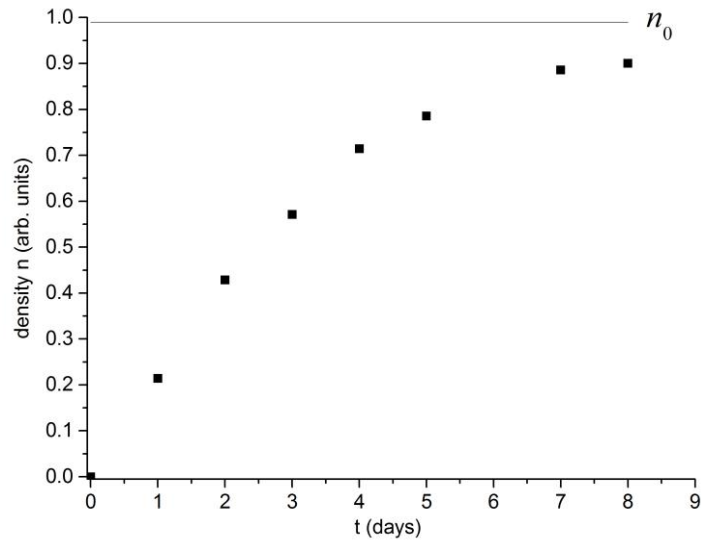

Fig. 2. Vapor density n versus curing time

The vapor density in the bulb may appear small due to the still existing trace of chemical active gases in the coating, which can remove the Rb atoms from the cell volume. We check if this possible mechanism is negligible by filling the semi-cured cell by atmospheric air for three days. Soon after, we pump the cell and start curing process again. We have found that in this case the density recovers to the previous level much faster, i.e. during several hours. It clearly demonstrates that the inserted chemical active impurities in the cell are completely pumped away in a relatively short time and the bonding of the Rb atoms by residual impurities in bulk of the coating of the properly pumped and cleaned cell is negligible.

We also check the possibility of the existence of chemical reaction of the Rb atoms with the PDMS coating. For that, we open and close the atomic source and record a variation of the residual - gas pressure (mainly hydrogen) in the cell. It is found that with the valve 5 being open the pressure of the gas is higher on several units of $10^{-10}$ mbar. This change of vacuum is partially due to a slight degassing of the atomic source, but also partially due to chemical reaction between the Rb and the PDMS. If each Rb atom, which is chemisorbed by the PDMS film, produces one hydrogen molecule, the loss of the Rb atoms density due to chemical bonding is about $10^{-10}$ mbar. This value is on three orders of magnitude less than working density of the Rb vapor at $20\ ^0C \sim 10^{-7}$ mbar [18].

From these results one can conclude that low saturated vapor density is mainly due to of the existence of the continuous loss of atoms via the adsorption of the atoms onto the glass substrate of the coating. The atoms from the source (where the atomic density is maximal) diffuse through the capillary to the cell vacuum volume. Then the atoms, after some bounces in the cell, become physadsorbed by the coating surface and then they start to diffuse deep inside the coating towards the glass substrate surface, where they are irreversibly trapped.

A flow of atoms in the capillary $J_c$ in Knudsen regime can be expressed as:

$$J_s = \frac{D_s S_s (n_0 - n)}{L_s} = \frac{2 r_s V_t}{3} \cdot \frac{\pi r_s^2}{L_s}(n_0 - n) = \frac{2\pi r_s^3}{3 L_s} V_t (n_0 - n), \quad (1)$$

where $D_s$ is diffusion coefficient of Rb atoms inside of capillary - $2 r_s V_t /3$; $S_s$ is its cross section - $\pi r_s^2$; $(n_0 - n)/L_s$ – density gradient; $V_t$ is thermal velocity $\sqrt{\frac{8kT}{\pi m}}$. Due to a complete bonding of the Rb atoms by the glass surface of the polymer substrate, the density n on the glass surface is equal to zero and the flow $J_p$ of the Rb atoms through the PDMS film can be written as:

$$J_p = \frac{D_p S_p n}{L_p} = \frac{D_p}{L_p} \cdot \left(4\pi R^2 + 2\pi R_V L_V\right) n \quad (2)$$

where $D_p$ is the diffusion coefficient of the Rb atoms in the coating bulk; $S_p$ is the sum of squares of the inner surfaces of both bulb and exit tube; $n/L_p$ – an atomic density gradient inside of the coating. We neglect the contribution of a relatively small area of the inner surface of the capillary and a low chemical loss of the atoms in the whole cell vacuum volume and in the coating. Thus, we can write that $J_c = J_p$ and calculate the diffusion coefficient $D_p$:

$$D_p \approx \frac{r_c^3 V_t}{3 L_p \left(2R^2 + R_V L_{V_c}^2\right)} \left(\frac{n_0}{n} - 1\right) \quad (3)$$

Figure 3 shows the diffusion coefficient $D_p$ of the Rb atoms in the coating bulk as a function of the curing time in the cell. One can see that $D_p$ when time equals zero is extremely large.

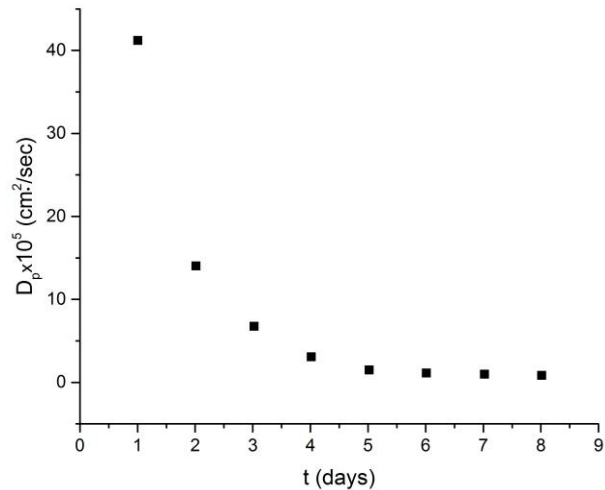

Fig. 3. Diffusion coefficient $D_p$ versus curing time

It decreases during the time of curing and reaches its low saturated value after about 8 days. One can see that value of $D_p$ after curing **is** decreased from $5·10^{-4}$ cm²/sec to $1.1·10^{-5}$ cm²/sec i.e. about 50 times.

It has to be noted that the permanent gradient of the density of the atoms from the atomic source ($n_0$) to the substrate surface (n = 0) causes a weak continuous flow of atoms. It takes a long time for the cell to achieve a truly steady state. In fact, for the capillary dimensions $r_c = 0.15$ см and , $L_c = 4.5$ см, Rb thermal velocity $2.7·10^4$ cm/s, density $n_0 = 5.2·10^9$ см$^{-3}$ [18] and density gradient after 8 days curing $(n_0 - n) / L_c = 1.156·10^8$, Eq. 1 yields flow = $2.2·10^{10}$ сек$^{-1}$ . For the cell inner walls square ~ 721 cm² it takes ten years to produce a monolayer on the surface of the glass substrate with $10^{16}$ atoms per cm².

We estimate how deep atoms can diffuse inside a fresh polymer film $l$ during a dwell time $\tau_d$. For instance, in the case of rubidium and PDMS, we have $D_p = 1.1·10^{-3}$ cm²/sec (that is found by extrapolation the curve in Fig.2 to zero time), $\tau_d = 10^{-4}$ sec from [19] and by using the formula $l = 2\sqrt{D_p \tau_d}$ we find $l = 9,4\mu$. This value is comparable with the thickness of coating of 10 μ used in this experiment.

The question arises – what has happened with the coating during curing process?

We performed another experiment for a more detailed understanding of the interactions between the Rb atoms and the PDMS. First, we measure the diffusion coefficient of Rb in bulk of a liquid PDMS compound which is free from ether. We use a setup consisted of a quartz cuvette with dimensions 10x10x40 mm, which is filled by the polymer, the source of the Rb atoms on the top of the cuvette, and a halogen lamp. The atomic source is a small rectangular box, made of a narrow stainless steel wire mesh filled by the PDMS with metallic Rb. Light from the lamp passed sequentially the aperture (diameter 1 mm), the cuvette and hits the spectrograph (AvaSpec-2048TEC), which detects the absorption of the light by the Rb atoms immersed in the PDMS. To eliminate the influence of a thermal convection in the polymer on the Rb diffusion, the cuvette was placed in a massive aluminum block with holes for the passage of light.

At the beginning of the measurements, we clean the PDMS of both the cuvette and the atomic source from chemical active gaseous impurities (particularly water). For that the cuvette and the box, both filled by the PDMS, are kept under vacuum for several days at a temperature of 110 $^0$ C. This period is typically sufficient to remove all chemical active gases from the polymer. Then we remove a piece of Rb from an ampoule and place it on the bottom of the box. To dissolve the metal, we place the box in vacuum for several days. This period is enough to dissolve the metal in the polymer and create a necessary high concentration of Rb in the source. After that, the box is tightly inserted into the upper part of the cuvette, until the flat bottom of the source (with the dissolved Rb) touches the surface of the polymer on the top of the cuvette. This allows the atoms to start to penetrate to the PDMS volume. This moment is taken as zero time (t = 0). The distance between the bottom of the box (x = 0) and the position of the light beam of the halogen lamp (x) inside the cuvette was chosen to be 10 mm, and it is measured with an accuracy of ±0.1 mm. Fig. 4 shows a typical transmittance spectrum of the Rb atoms which diffuse from the atomic source to the detection point with the halogen light beam.

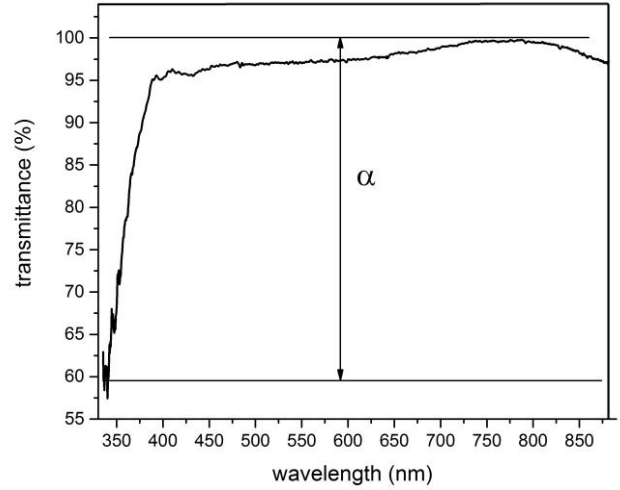

Fig. 4. Transmittance spectrum of the Rb atoms in the PDMS,
(spectrum of pure PDMS is subtracted).

hours. The figure shows that the spectrum is represented by two structures less absorption bands: the first band extending from 350 nm to 400 nm; the second - from 780 nm to 880 nm.

The concentration of the Rb atoms C is defined by a standard diffusion equation:

$$D\frac{d^2C}{dx^2} = \frac{dC}{dt} \quad (1)$$

with an initial condition $C(x) = 0$ for $t = 0$ and boundary conditions $C(x) = C_0$ at $x = 0$; $C_0$ - concentration of the atoms in the source, which is assumed to be a constant. The solution of the equation 1 is the following:

$$C = C_0[1-\Phi(z)], \quad \Phi(z) = \frac{2}{\sqrt{\pi}}\int_0^z \exp(-y^2)dy, \quad z = \frac{x}{2\sqrt{Dt}}. \quad (2)$$

where D - diffusion coefficient. The diffusion coefficient of Rb is calculated from two measurements of the concentrations $C_1$ and $C_2$ which corresponds to two different times $t_1$ and $t_2$. To exclude an unknown constant $C_0$ we use ratio of $C_1$ and $C_2$. According to the Beer's law, the atoms concentration C at the detection point is proportional to ln (100-α) /100, where α is the amplitude of the first largest band of the transmitted spectra in percents.

$$\frac{\ln(100-\alpha_1)/100)}{\ln(100-\alpha_2)/100)} = \frac{C_1}{C_2} = \frac{1-\Phi(z_1)}{1-\Phi(z_2)}, \quad z_1 = \frac{x}{2\sqrt{Dt_1}}, \quad z_2 = \frac{x}{2\sqrt{Dt_2}}. \quad (3)$$

The value of the diffusion coefficient is obtained by averaging of 5 results of the measurements. All calculations are done by the adoption of $D_p$, which satisfies the formula (3). The average value of the diffusion coefficient of rubidium atoms was found to be equal to 0,99 · $10^{-5}$ cm$^2$ / sec. The statistical uncertainty is ± 5%, which is due to a small non-compensated thermal convection of the PDMS in the cuvette. Both coefficients measured in the cell - 1.1 •$10^{-5}$ cm$^2$/sec and in the cuvette - 0,99 · $10^{-5}$ cm$^2$ / sec - are equal within error bar measured in [20] - (1,2 ± 0,7) · $10^{-5}$ cm$^2$ / sec.

Second, we measured the diffusion coefficient again in the used polymer but we had little success. For that we took the used cuvette which was kept under the vacuum for three months, removed the old atomic source and put inside the cuvette a new one. In order to detect a new Rb absorption spectrum on the background of the old one, the temperature of the new source was increased up to 40 $^0$C. We found it very difficult to detect an extra absorption of light which was produced by a higher density of Rb at the detection point and measure the diffusion coefficient. This is due to both a small transmittance of light of the used PDMS and an extremely slow diffusion of the Rb atoms in it. We extracted the used PDMS from the cuvette and found that the compound looks like a piece of rubber: it is white colored, not transparent and rigid.

It is known that the PDMS compound has several physical and chemical properties: a unique flexibility, low temperature variations of the physical constants, high dielectric strength, extremely low chemical reactivity with alkali atoms, low glass transition and melting temperatures and a high gas permeability and diffusivity (see, for example [21] and references therein).

We performed the measurements of some of these parameters of the PDMS treated by Rb atoms by using by a differential scanning calorimeter (DSC 200 F3 Maia, Netzch. We measured the PDMS glass transition temperature $Tg$ of the polymer. We found that $Tg$ for the PDMS is equal to – 70 ◦C. This value of the glass transition temperature is higher than $Tg$ previously measured in liquid polymer (– 125 $^0$C). The melting temperature $Tm$ was found to be equal to 70 $^0$C which is again higher than that temperature of those of none treated PDMS (- 40 $^0$C). We failed to measure the viscosity of this practically solid polymer, but the estimations show us that the viscosity has been enhanced up to about million times: from 4•10$^3$ cm$^2$s$^{-1}$ up to about 10$^9$ -10$^{10}$ cm$^2$s$^{-1}$. Note, that these measured parameters have a profound influence on polymer's transport properties: typically, a polymer with high glass transition and melting temperature and high viscosity possesses low diffusivity [22]. They are roughly similar to the parameters of sodium - catalytic polymerized isoprene rubber [23]. To our knowledge, the polymerization of the PDMS under the catalytic action of the Rb atoms has been observed by us for the first time [24].

Based on the experimental findings, one is led to interpret the observed dependence of the diffusion coefficient on the curing time. Initially, soon after its preparation, the PDMS coating is a porous film, with a high permeability and diffusivity. The atoms, inside the cell vacuum volume, rapidly diffuse through the coating towards the glass substrate, where they become irreversibly trapped. This causes a low vapor density in the cell. During the curing process, cross-links are added by the catalytic action of the Rb atoms to long PDMS molecules. When polymer chains are linked together by cross-links, they lose some of their ability to move as individual polymer chains. The high cross-link density decreases the viscosity of the PDMS, free volume and it transforms the compound to a rigid material with a low porosity. As a result, after curing the PDMS film has a significantly smaller diffusion coefficient which means that an impeding barrier is formed in the coating. It leads to a high vapor density in the cell. Since the atoms do not diffuse inside the coating, the dwell time should be shorter than the one on a fresh coating. According to the Stokes – Einstein equation, which describes diffusivity to be inverse proportional to viscosity, one can speculate that on the surface of a completely polymerized PDMS with viscosity enhanced one million times, the dwell time should be one million times smaller: 10$^{-4}$/10$^6$ = 10$^{-10}$ s. This is true if the polymerization does not change the attractive potential barrier on the cured PDMS film surface. Note that this estimated dwell time value is close to the one derived from the measured potential barrier on the film surface and the film temperature [19].

**Acknowledgements**

We would like to thank Dr. V. Surovtsev and Dr. A. Pugachev for stimulating discussions. Special thanks for R. Robson (McKillop) for careful reading of the manuscript. The work is supported by Russian Foundation for Basic Research (RFBR), grant number 15-02-02333 A